\documentstyle[prl,aps,twocolumn,epsf]{revtex} 
\newcommand{\be}{\begin{equation}} 
\newcommand{\ee}{\end{equation}} 
\newcommand{\bea}{\begin{eqnarray}} 
\newcommand{\eea}{\end{eqnarray}}

\begin{document} 

\twocolumn[\hsize\textwidth\columnwidth\hsize\csname @twocolumnfalse\endcsname 

\title {Condensation energy in the spin-fermion model for cuprates} 
\author{Ar. Abanov and  Andrey V. Chubukov} 
\address{ 
Department of Physics, University of Wisconsin, Madison, WI 53706} 
\date{\today} 
\draft 
\maketitle  
\begin{abstract} 
We use the Scalapino-White relation between the condensation 
energy and  the difference between the dynamical structure
factor in the normal and the superconducting states to
compute the condensation energy
in the spin-fermion model. We show
that at strong coupling, the extra low-frequency 
spectral weight associated with the resonance peak in the dynamical structure
factor in a superconductor is compensated only 
at energies $\sim J$ which are much
 larger than the superconducting gap $\Delta$. We argue that in this situation,
the condensation energy 
is large and well accounts for the data for cuprates. 

\end{abstract} 

]
 \narrowtext 

The understanding of the mechanism of superconductivity is an important step
towards the general understanding of the physics of cuprates.
It has been know from the studies of BCS superconductors, that 
the information
about the pairing boson can be extracted from measurements of the upper
critical field. Specifically, the
increase of the kinetic energy in a superconductor is overcompensated by the
decrease of  the potential
energy associated with the feedback effect from superconductivity 
on the bosonic mode which is
responsible for pairing~\cite{schrieffer}. The energy difference is called a
condensation energy $E_c$ and is 
 directly related to the measurable 
thermodynamic  critical field by $E_c = V_0~H^2_c/(8\pi)$  where $V_0$ is the
volume of the unit cell. 

Recently, Scalapino and White~\cite{sw}  applied this reasoning to high $T_c$
superconductors. They argued that 
if the pairing is mediated by spin fluctuations, then 
the difference in the dynamical structure factor 
$S (q, \Omega)$
between the  normal and the 
superconducting states, integrated over frequency and 
momentum with the weighting factor $(\cos q_x +\cos q_y)$ should
be {\it positive} and of the same order as $E_c$. 
This yields a relation~\cite{sw,demler2}
\begin{eqnarray}
\frac{H^2_c}{8\pi} &=& \frac{3 N}{2} \alpha J \int \frac{d^2
q}{4\pi^2}~\int_0^{\infty} \frac{d \Omega}{\pi} \nonumber \\
&&\times\left(S_n (q,\Omega) - S_{sc} (q,\Omega)\right)~(\cos q_x + \cos q_y),
\label{rel}
\end{eqnarray}
where $\alpha<1$ is a numerical factor which accounts for the fact that
the condensation energy is smaller than the decrease in the potential energy,
  and $N$ ($=2$ for $YBCO$ and $Bi2212$) 
is the number of layers in the unit cell. 

Neutron scattering experiments in $YBCO$ and $Bi2212$ 
demonstrated~\cite{fong,neutrons}   that
in a superconducting state, $S_{sc} (q,\Omega)$ possesses a
resonance peak at  momenta near $Q =(\pi,\pi)$ and at  frequencies below 
$2\Delta$ where $\Delta \ll J$ is the 
maximum of the $d-$wave gap. 
The integrated intensity of the resonance peak yields the 
r.h.s. of (\ref{rel}) consistent with the data on $H_c$.  
However, it is not clear a'priori to which extent the
contribution from the resonance peak measures  the
actual condensation energy in a system. The relevant issue here is
 whether the spectral weight
of the resonance 
peak is compensated by the depletion 
of the spectral weight
in $S_{sq} (q,\omega)$ at energies comparable to $\Delta$ and hence at typical
$|q-Q| \sim \Delta/J \ll 1$, or the compensation comes from energies
comparable to $J$, i.e., from $|q-Q|$ of order 1. 

In the first case, the
geometrical $\cos q_x + \cos q_y$ factor is nearly constant for relevant $q$
and can be omitted.
Since $\int d^2 q d\Omega S (q,\Omega) = S(S+1)/3$ where $S$ is the average
value of the on-site spin (this is the sum rule for spin structure factor),
  the r.h.s of (\ref{rel}) then just
measures the difference in $S$ between normal and superconducting states. 
In general, this difference is finite due to a possibility for double occupancy
for which case $d-$wave superconductivity favors a spin singlet, $S=0$ state.
However, in cuprates double occupancy is energetically
unfavorable (Hubbard $U$ is large), and the condensation energy
should be small. 

In the second case, however, typical $q$ are far from $Q$,
and the momentum dependence of the geometrical factor cannot be neglected. 
In this situation, one can expect that the condensation energy is not
reduced by the sum rule constraint and, generally, is of the same
order as the unscreened contribution from the resonance peak.  

There are two qualitatively different explanations of the resonance peak.
One was presented by us~\cite{ac}, and is based on 
strong coupling calculations within the spin-fermion model. These calculations 
extent earlier weak-coupling
results by others~\cite{others}.   We argued that the
resonance peak is in the  particle-hole channel  and is 
related to the fact that 
in a superconductor, the damping of a spin fluctuation due to a decay into
a particle-hole pair is strong only at frequencies above $2\Delta$, while 
below $2\Delta$ it is strongly reduced because of a lack of 
phase space for a decay. By a 
Kramers-Kronig relation, this reduction produces a real part of the spin
polarization bubble. At low energies, this real part scales as $\omega^2$, i.e
spin collective modes in a superconductor behave as propagating magnons.
 This behavior obviously 
gives rise to a peak in $S_{sc} (q, \Omega)$ at
$\Omega=\Omega_{res} \propto \xi^{-1}$, where $\xi$ is 
the magnetic correlation length. 

Another explanation was presented by 
Demler and Zhang~\cite{demler} in the context of $SO(5)$
theory of superconductivity. They  conjectured that the peak seen in neutron
scattering is an antibound state in the spin-triplet, 
particle-particle channel at total momentum $Q$ ($\pi$ resonance).  Below $T_c$,
particle-particle and particle-hole channels are mixed, and the antibound state
appears as a pole in the spin susceptibility.

Demler and Zhang recently argued~\cite{demler2}
 that the measurement of the 
condensation energy is a way to distinguish between the two theories.
They conjectured on general grounds
 that if the $\pi$ resonance is the correct explanation, then it is likely that
the compensation of the peak spectral weight comes from high energies. 
Their argumentation is that since 
the antibound state is by itself insensitive to $T_c$, it is energetically
favorable for a system to undergo a superconducting transition so that 
a $\pi-$resonance can emerge and increase the condensation energy.
They also argued that in the spin-fluctuation theory of the peak, the
compensation of the spectral weight is confined to a vicinity of $2\Delta$, and
hence the condensation energy is small.

In the present paper we show that this is not the case. We compute $S(q,\Omega)$ in the spin-fermion model and show that 
the compensation of the  spectral weight associated with the resonance peak 
in fact comes from high energies $\sim J$ or, equivalently, from momenta $q$ 
far from $Q$. We also argue that at strong
coupling, the low-energy 
antibound state in the particle-particle channel does not exist
because fermionic incoherence washes out the upper boundary of fermionic dispersion. 

The point of departure for our consideration is the spin-fermion model for
cuprates. It describes low-energy fermions interacting with their 
collective spin 
degrees of freedom~\cite{pines,chubukov}. 
Of interest here is the form of the full dynamical spin susceptibility.
It has been derived in earlier studies~\cite{ac,chubukov}, and we just
quote the results. Both in the normal and the superconducting state,
the full spin susceptibility can be written as 
$\chi^{-1} ({\bf q}, \Omega) = \chi^{-1}_0 ({\bf q}) - 
\Pi_{{\bf q}} (\Omega)$, where 
 $\chi_0$ is the bare susceptibility which is
made of fermions with energies comparable to bandwidth, and 
$\Pi_{\bf q} (\Omega)$ is the universal (i.e. cutoff independent) 
contribution from low-energy fermions.

The form of $\chi_0$ is the input for low-energy calculations.
 As before~\cite{ac,chubukov}, we assume 
that $\chi_0$ is peaked at $Q$ or near $Q$, and has a
 simple Ornstein-Zernike form i.e.
$\chi_0 ({\bf q}) = \chi_0 \xi^{2}/(1 + ({\bf q} - {\bf Q})^2 \xi^2)$. 

The universal contribution to the dynamical 
susceptibility involves low-energy fermions and therefore has to
be computed fully self-consistently within the spin-fermion model. 
Near $q =Q$, one can neglect $q$ dependence in $\Pi$ (it yields only a small
correction to already excising dispersion in $\chi_0 ({\bf q})$) and  
restrict with $\Pi_Q (\Omega) = \Pi_\Omega$. 
The full susceptibility then has the form
\begin{equation}
\chi ({\bf q}, \Omega) = \frac{\chi_0 \xi^{2}}{1 + 
({\bf q} - {\bf Q})^2 \xi^2 -
\Pi_\Omega}.
\label{ch}
\end{equation}
We absorbed $\chi_0 \xi^2$ factor into the redefinition of
$\Pi_\Omega$.
                                                           
In the normal state,  $\Pi_\Omega$
 is purely imaginary and for any coupling strength is
almost linear in $\Omega$:
$\Pi_\Omega =i |\Omega|/\omega_{sf}$ 
where $\omega_{sf} \propto \xi^{-2}$.
The deviations from the linear behavior result from 
 the corrections to the particle-hole vertex which at $\xi = \infty$ are
logarithmical in $\omega$. However, the prefactors are small, and the
deviations from linearity become relevant only in the extremely tiny region
near $\xi = \infty$ which we will not study here. 

Consider now the superconducting state. 
Here the form of $\Pi_\Omega$ is more complex
because spin damping is cut below $2\Delta$~\cite{ac,others}. 
By Kramers-Kronig relation, this cut in  $Im\Pi_\Omega$ 
creates $Re \Pi_\Omega$ which, as we said before, gives rise to a
resonance peak in $\chi^{\prime\prime} (Q,\Omega)$ below $2\Delta$. 

In general, the spin polarization operator 
in a superconductor is a sum of bubbles made of
normal and anomalous Green's functions. Both fermions in the bubble has to be
near the Fermi surface to satisfy the constraint on energy conservation. For
$q\approx Q$, this restricts the momentum integration to the vicinity of hot
spots- points at the Fermi surface separated by $Q$. 
As in ~\cite{ac}, we
consider the situation near optimal doping (when pseudogap effects are
weak) and approximate the gap near hot spots by a frequency independent
input parameter $\Delta$. Under this
approximation, 
the fermionic self-energy $\Sigma_\omega$ and  
the spin polarization operator $\Pi_\Omega$ are given by a set
of two coupled equations~\cite{ac}:
\begin{equation}
\Sigma_\omega = \omega + \frac{3R}{8 \pi^{2}}~\int
\frac{\Sigma_{\omega +\Omega}}
{q_{x}^{2}+\Sigma ^{2}_{\omega +\Omega}-\Delta ^{2}}
\frac{d \Omega d q_{x}}{\sqrt{q_{x}^{2}+1-\Pi_\Omega }} 
\label{set2a}
\end{equation}
\begin{equation}
\Pi_\Omega= \frac{i}{2}
\int \frac{d \omega}{\omega_{sf}}~\left(           
\frac{\Sigma_{\Omega -\omega}~\Sigma_\omega + \Delta ^{2}}
{\sqrt{\Sigma^{2}_{\Omega -\omega}-\Delta ^{2}}~
\sqrt{\Sigma^{2}_\omega-\Delta ^{2}}}+1 \right).
\label{set2b}
\end{equation}
Here $R = {\bar g}/(v_{F} \xi^{-1})$ 
is a dimensionless parameter which governs the strength of the
spin-fermion coupling (we use the same notations as in ~\cite{ac} - ${\bar g}$
is the effective spin-fermion coupling, 
$v_{F}$ is the Fermi velocity at a hot spot). 
There are numerous reasons to believe that at 
and below optimal doping $R\gg 1$. 
To shorten notations, we included a bare $\omega$
term in $G^{-1} (k, \omega)$ into the self-energy.

We discuss the solution of (\ref{set2a},\ref{set2b}) 
below but first 
consider what we actually need to compute. 
Our goal is to check how the extra spectral
weight in local $S_{sc} (\Omega)$
 is redistributed compared to the normal state. For this
purpose, it is sufficient to compute the integral in
(\ref{rel}) without the geometrical $\cos q_x + \cos q_y$ factor and just
check at which scales the sum rule is recovered.

Without $\cos q_x + \cos q_y$, the momentum integration in the r.h.s. in
(\ref{rel}) can be performed exactly, and at $T\rightarrow 0$, we obtain using
$S (q, \Omega) = 2
\chi^{\prime \prime} (q, \Omega) /(1 -e^{-\hbar \Omega/T})$ 
\begin{equation}
I = \int \frac{d^2 q d \Omega}{4 \pi^3}~\left(S_{sc}
(q,\Omega) - S_n (q,\Omega) \right)  =\frac{\chi_0}{4 \pi^2} 
\int_0^\infty d \Omega~F (\Omega),
\label{chio1}
\end{equation}
where
\begin{equation}
F(\Omega) = \arctan 
\frac{\omega_{sf}}{\Omega} - \arctan\frac{1-Re \Pi_\Omega}{Im \Pi_
\Omega}.
\label{chio}
\end{equation}
We see that the rate of convergence of  the r.h.s. of (\ref{chio1}) depends
on the forms of both $Re \Pi_\Omega$ and $Im \Pi_\Omega$ above
the superconducting gap.

We now  obtain these forms from Eq. (\ref{set2b}). 
Qualitatively,  the solution  of this set has been 
obtained earlier~\cite{ac,chubukov}. Here we present  quantitative
results for $\Pi_\Omega$.

At $R \gg 1$, the normal state self-energy has a Fermi liquid form
$\Sigma (\omega) \propto Z^{-1}(\omega + i \omega |\omega|/(4 \omega_{sf}))$
with $Z \propto R^{-1} \sim
(\omega_{sf}/{\bar g})^{1/2}$ at energies smaller than, 
and at larger frequencies 
crosses over into a non-Fermi
liquid, quantum-critical regime $\Sigma (\omega) \propto \exp{(i\pi/4)}~
\omega~\sqrt{\bar {g}/|\omega|}$. A simple experimentation shows that
 the solution of 
(\ref{set2a},\ref{set2b})
depends on the ratio between $\omega_{sf}$ and the measured superconducting gap
which is ${\bar \Delta} = \Delta Z$ if $\omega_{sf} \gg {\bar \Delta}$, and
 ${\tilde \Delta} \sim\Delta^2/{\bar g}$ if
$\omega_{sf} \ll {\bar \Delta} \ll {\tilde \Delta}$. In the first case, 
at typical frequencies $\sim {\bar \Delta}$
 the system behaves in the normal state as a Fermi-liquid, while in the second
case, which is more relevant to optimally doped and underdoped
cuprates~\cite{chubukov}, fermions with $\omega \sim {\tilde 
\Delta} \gg \omega_{sf}$ display in the normal state 
the quantum-critical, $\sqrt{\omega}$ behavior.
 \begin{figure}
  \centerline{\epsfxsize=3.0in \epsfysize=2.0in 
  \epsffile{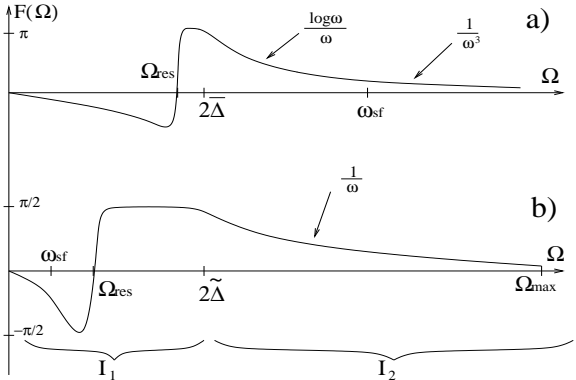}}
  \caption{Schematic representation of $F(\Omega)$ from Eq. 
~\protect(\ref{chio1},\ref{chio}) for (a)  weak coupling case 
$\omega _{sf}\gg {\bar\Delta}$ and 
(b) strong coupling case 
$\omega _{sf}\ll {\bar \Delta} \ll{\tilde \Delta}$. Here 
${\bar \Delta}$ and ${\tilde \Delta}$ are measured superconducting 
gaps at weak and strong coupling, respectively, 
$\Omega_{res}$ is the  frequency of the neutron resonance peak in a superconductor, and $\omega_{sf}$ is a typical spin relaxation frequency in the normal state. Observe that in both cases, 
the frequency integral of $F(\Omega)$ is confined to frequencies, which are much larger than the measured superconducting gap.}
\label{fig1}      
\end{figure}
We now consider the two cases separately. At
$\omega_{sf} \gg {\bar \Delta}$, the quasiparticle residue 
just renormalizes the superconducting gap ($\Delta \rightarrow {\bar \Delta}$),
and the system behavior is the same as 
at weak coupling. In this situation~\cite{others}
$Im \Pi_\Omega =0$ for 
 $\Omega < 2{\bar \Delta}$,
 while $1-Re \Pi_\Omega$ changes sign at
$\Omega_{res} = 2 {\bar \Delta} (1 - O(e^{-\omega_{sf}/{\bar \Delta}})$.
Above  $2{\bar \Delta}$, the analytical form for $\Pi_\Omega$ can
be obtained in the limit of $\Omega \gg 2{\bar \Delta}$. We found
$Re \Pi_Q(\Omega) \approx \pi {\bar \Delta}^2/(\Omega \omega_{sf})$ and
$Im \Pi_Q(\Omega) = (\Omega/\omega_{sf}) + (2{\bar \Delta}^2/(\Omega
\omega_{sf}))~\log (\Omega/{\bar \Delta})$. 
Substituting these results into (\ref{chio}) we find after a simple algebra
that below $2{\bar \Delta},~ F(\Omega)$ is negative except for a tiny range
between $\Omega_{res}$ and $2{\bar \Delta}$, while above $2{\bar \Delta}$,
$F(\Omega)$ is positive and scales as
$F(\Omega) \propto (1/\Omega)~\log \Omega/{\bar \Delta}$ for $\Omega <
\omega_{sf}$ and as $F(\Omega) \propto (1/\Omega)^3$ for $\Omega >
\omega_{sf}$. This behavior is schematically shown in Fig. (\ref{fig1}a).
Splitting the integral in (\ref{chio1}) 
in two parts, $I=I_1 + I_2$, where
 the first is the integral over frequencies up to twice the
measured gap and the second is the integral over larger frequencies, and
performing integration we obtain with the logarithmical accuracy
\begin{eqnarray}
I_1 &\approx& \frac{\chi_0}{4\pi^2}~\left(\pi (2{\bar \Delta} - \Omega_{res}) -
\frac{2{\bar \Delta}^2}{\omega_{sf}}\right) \approx
-\frac{\chi_0}{2\pi^2}~\frac{{\bar \Delta}^2}{\omega_{sf}} \nonumber \\
I_2 &\approx& \frac{\chi_0}{2\pi^2}~\frac{{\bar \Delta}^2}{\omega_{sf}}
~\int_{\sim {\bar \Delta}}^{\sim \omega_{sf}} ~\frac{d
\Omega}{\Omega}~\log{\frac{\Omega}{\bar \Delta}}
\approx  \frac{\chi_0}{4\pi^2}~\frac{{\bar \Delta}^2}{\omega_{sf}}
\log^2{\frac{\omega_{sf}}{\bar \Delta}} 
\label{weak}  
\end{eqnarray}
We see that the contribution from low frequencies is negative - the vanishing
of $Im \Pi_\Omega$ overshadows the contribution from the resonance peak
which for $\omega_{sf} \gg {\bar \Delta}$ has an exponentially small residue.
However, the contribution from frequencies above $2{\bar \Delta}$ is positive
and parametrically 
larger than negative $I_1$, such that $I = I_1 + I_2 >0$.

It is essential that although typical frequencies in $I_2$
are $\Omega \sim \omega_{sf} \gg {\bar \Delta}$,
still $I_2$ converges at 
$\omega > \omega_{sf}$ and therefore does not depend on system behavior 
far away from $Q$. Indeed, at $\omega \sim \omega_{sf}$,
 typical $q-Q$ are of the order of inverse
correlation length (see (\ref{ch})). 
 The geometrical $\cos q_x + \cos q_y$
factor is then nearly a constant ($=-2$) for all relevant $q$ and hence a 
nonzero value
of the condensation energy just reflects the fact that in the absence of no
double occupancy constraint, the average on-site spin in the $d-$wave 
superconducting
state is smaller than in the normal state.
Notice  that the magnitude of $I$ is much smaller
than $\chi_0 {\bar \Delta}$ which would be  the
contribution from the resonance peak if its residue  was $O(1)$

Consider now the opposite case of 
 $\omega_{sf} \ll {\bar \Delta} \ll {\tilde \Delta}$ where, we remind,
${\tilde \Delta} \sim \Delta ^{2}/{\bar{g}}\sim {\bar \Delta}^2/\omega_{sf}$ 
is the measured gap
in this limit. Below $2{\tilde \Delta}$ we still have $Im~ \Pi_\Omega =0$,
but the resonance frequency (the one at which $Re~ \Pi_\Omega =1$)
is now much smaller than the gap:
 $\Omega_{res} \sim ({\tilde \Delta} \omega_{sf})^{1/2} \ll 
{\tilde \Delta}$~\cite{ac}.
 At $\Omega \gg {\bar \Delta}$ but still $\Omega \ll {\bar
g}$, we found from (\ref{set2a}, \ref{set2b}) 
that $Im ~\Pi_\Omega$ approaches the normal state form 
$|\Omega|/\omega_{sf}$, but $Re~ \Pi_\Omega$ saturates at
$Re~ \Pi_\Omega  = (\pi {\tilde \Delta}/(2\omega_{sf}) (1 + \pi^{-1} \log 4)$
and preserves this value as long as 
the fermionic propagator has a non-Fermi liquid, $\sqrt{\omega}$
form. At very large frequencies 
 $Re \Pi_\Omega $ indeed decreases, but the decrease begins only at 
$\Omega \sim \Omega_{max}$, 
where either the fermionic propagator recovers Fermi-liquid
behavior, i.e., bare $\omega$ term exceeds $\sqrt{\omega}$ contribution from
the self-energy, or typical $q-Q$ in $S(q,\Omega)$ become $O(1)$, i.e., lattice
effects become relevant. For ${\bar g} \ll v_F k_F$, the recovery of the
Fermi-liquid behavior comes first, and
$\Omega_{max} \sim {\bar g}$. For ${\bar g} \gg v_F k_F$ 
(which in Hubbard-model language
implies $U \gg t$), lattice effects become relevant first, 
and $\Omega_{max} \sim (v_F k_F)^2/{\bar g} \sim J$. 

Substituting the results for $\Pi_\Omega$ into (\ref{chio}) we find that
$F(\Omega)$ is now positive at frequencies below $2{\tilde\Delta}$, 
except for very low $\Omega < \Omega_{res} \ll 2{\tilde \Delta}$. However,
$F(\Omega)$ is also positive above $2{\tilde \Delta}$ and, moreover,
due to a saturation in $Re \Pi_\Omega$, it
 behaves as $F(\Omega) \propto 1/\Omega$  up to $\Omega_{max}$ which again
causes a logarithmical behavior of the frequency integral. This behavior of
$F(\Omega)$ is schematically shown in Fig. (\ref{fig1}b).
Evaluating the integral in (\ref{chio1}), 
with the logarithmical accuracy we obtained
\begin{eqnarray}
I_1 &=& \frac{\chi_0}{4\pi}~({\tilde \Delta} - \Omega_{res}) \approx
\frac{\chi_0}{4\pi}~{\tilde \Delta} \nonumber \\
I_2 &=& \frac{\chi_0}{8\pi}~ {\tilde \Delta} \beta~ \int_{\sim {\tilde
\Delta}}^{\Omega_{max}} \frac{d \Omega}{\Omega} = \frac{\chi_0}{8\pi}~
{\tilde \Delta} \beta~ \log{\frac{\Omega_{max}}{\tilde \Delta}}
\label{str}
\end{eqnarray}
where $\beta = 1 + \pi^{-1} \log 4$. 
We see that $I_1$ is positive, i.e., the
 appearance of the
resonance peak below $T_c$ gives rise to an extra integrated 
spectral weight below $2{\tilde \Delta}$ and hence yields  a positive
contribution to the condensation energy. As we discussed before, this 
extra spectral weight should be compensated by a depletion of the spectral
weight at somewhat higher frequencies. We see however that due to non-Fermi
liquid behavior of the fermionic propagator above $2{\tilde \Delta}$, this
compensation comes from frequencies larger than $\Omega_{max}$.
Moreover, the integrated 
contribution from energies between $2{\tilde \Delta}$ and
$\Omega_{max}$ is logarithmically  
larger than the contribution from the resonance peak.
Further, for ${\bar g} \gg v_F k_F$ (the case when the no double occupancy
constraint is almost exact), 
$\Omega_{max} \sim \omega_{sf} \xi^2$ and hence
 typical momenta in $\chi (q,
\Omega)$ are $(q-Q)^2 \xi^2 \sim \Omega_{max}/\omega_{sf} \sim \xi^2$, or 
$q-Q = O(1)$. In this situation, the geometrical
$\cos q_x + \cos q_y$ factor in (\ref{rel}) cannot be approximated by a
constant and effectively reduces the contribution from high energies. In other
words, even if $\int d^2 q d \Omega S(q,\Omega)$ does not change between normal
and superconducting states, 
{\it there is still a finite, positive condensation energy which 
is even larger than
the net contribution from the resonance peak}. This is
the central result of the paper. 

The magnitude of the 
contribution to $E_c$ from the  resonance peak in
optimally doped $YBCO$
has been estimated in~\cite{demler2} without invoking any theory but 
rather using the experimental
results for $\int \chi^{\prime \prime} (q,\Omega)$~\cite{fong}. They found
$E_c \sim 0.03 \alpha J$ where, we
recall, $\alpha <1$ accounts for the reduction of $E_c$ due to the increase in
the kinetic energy. Using $J \sim 1500$, one 
obtains $E_c \sim 45 \alpha K$ which, as Demler and Zhang argued~\cite{demler2}
agrees with $E_c \sim 3-12K$ extracted from penetration depth and
specific heat measurements.
Our results show that the actual magnitude of $E_c$ is
even higher due to an extra positive contribution from frequencies above
$2{\tilde \Delta}$. Furthermore, we found 
that this extra contribution to the condensation energy 
is  larger than the one from the peak. 
From this perspective, the above estimate 
for $E_c$ yields a lower boundary for the condensation energy. 

To summarize, in this paper we considered the condensation energy within the
spin-fermion model for cuprates. At strong coupling, this model
predicts that in a superconducting state, $\chi^{\prime \prime} (Q,\Omega)$
 possesses a sharp resonance peak below 
twice the maximum of the measured $d-$wave gap. 
We demonstrated that the appearance of this peak does not
cause the depletion of the spectral weight 
in local $\chi^{\prime \prime}$ 
up to frequencies of order $J$. 
We computed the condensation energy $E_c$ 
using Scalapino-White relation between $E_c$ and $\chi^{\prime \prime}$ and
found that the dominant, positive contribution to $E_c$ comes from a wide range
of frequencies between $2{\bar \Delta}$ and $J$. 
Our results disagree with the assertion in ~\cite{demler2} that a large
condensation energy cannot be obtained in the spin-fermion model and therefore
would require a resonance in the
triplet particle-particle channel ($\pi-$resonance). 

We on the contrary didn't find any indication of a sharp resonance in the 
$\pi$ channel at strong coupling. This resonance can only emerge as
 an antibound state and requires a sharp upper boundary of the fermionic
spectrum. We, however, 
 found that strong fermionic self-energy transforms
the spectral weight from the quasiparticle peak to higher frequencies and
washes out a sharp upper boundary of fermionic excitations. We caution however,
that our analysis is valid for a Fermi surface with hot spots. Without hot
spots, the fermionic decay is forbidden, and at least in
some range of couplings, the fermionic spectrum preserves
a sharp upper boundary in which case the system possesses an
antibound state in the $\pi$ channel.   

It is our pleasure to thank D. Basov, G. Blumberg, J.C. Campuzano,
E. Demler, M. Norman,  D. Pines, J. Schmalian, and S-C Zhang for useful 
conversations. The research was supported by NSF DMR-9629839.

\end{document}